\documentclass[aps,prl,twocolumn,superscriptaddress,raggedbottom,showpacs,floatfix]{revtex4-1}
\usepackage{amsmath,amsfonts,amssymb,amsthm,graphics,graphicx,latexsym,color}
\usepackage[next]{inputenc}
\usepackage[dvips]{epsfig}
\usepackage[colorlinks=true,citecolor=blue,linkcolor=blue]{hyperref}
\usepackage{bbm,bm, bbold}
\usepackage{booktabs}
\usepackage{multirow}
\usepackage{hhline}
\usepackage{comment}
\usepackage{float}

\usepackage{epstopdf}

\begin{document} 
\renewcommand{\vec}{\mathbf}
\renewcommand{\Re}{\mathop{\mathrm{Re}}\nolimits}
\renewcommand{\Im}{\mathop{\mathrm{Im}}\nolimits}

\title{Many-Body Theory of Trion Absorption Features in Two-Dimensional Semiconductors}
\author{Dmitry K. Efimkin}
\affiliation{The Center for Complex Quantum Systems, The University of Texas at Austin, Austin, Texas 78712-1192, USA}

\author{Allan H. MacDonald}
\affiliation{The Center for Complex Quantum Systems, The University of Texas at Austin, Austin, Texas 78712-1192, USA}

\begin{abstract}
Recent optical studies of monolayer transition metal dechalcogenides have demonstrated 
that their excitonic absorption feature splits into two widely separated peaks at finite carrier densities.  
The additional peak is usually attributed to the presence of trions, bound states of two electrons and a hole or an electron and two holes.
Here we argue that in the density range over which the trion peak is well resolved, it cannot be interpreted 
in terms of weakly coupled three-body systems, and that the appropriate picture is instead one 
in which excitons are dressed by interactions with a Fermi sea of excess carriers.
This coupling splits the exciton spectrum into a lower energy attractive exciton-polaron 
branch, normally identified as a trion branch, and a higher energy repulsive exciton-polaron branch, normally
identified as an exciton branch. 
We have calculated the frequency and doping dependence of the optical conductivity and found that: $\hbox{(i)}$ the splitting varies linearly with the Fermi energy of the excess quasiparticles; $\hbox{(ii)}$ the trion peak is 
dominant at high carrier densities; $\hbox{(iii)}$ and the trion peak width is considerably smaller than that of the excitonic peak. 
Our results are in good agreement with recent experiments. 
 \end{abstract}
\maketitle

\noindent
\section{I. Introduction} A decade ago  graphene introduced two-dimensional massless Dirac fermions to condensed matter physics~\cite{Graphene1,Graphene2,GrapheneMacDonald,GrapheneRMP}. Graphene was the first member of a large and still growing 
family of {\it flatland} materials, which includes the two-dimensional transition metal dichalcogenides (TMDCs) ~\cite{TMDC1,TMDC2,TMDC4,TMDC5,TMDC6}.  Monolayer TMDCs exhibit exceptionally strong spin-orbit and 
electron-electron interaction effects, and for this reason have provided a rich new playground for the exploration of 
exciton physics.  
TMDC excitons have strong excitonic absorption features with large binding energies ($\sim 0.5~\hbox{eV}$)
that dominate the optical absorption properties addressed in this paper (See Ref.~\cite{TMDCReview} for a review).  

An important feature of two-dimensional semiconductors is the possibilities they offer for 
controlling optics by gating.  Recent experiments~\cite{TrionExperiment1,TrionExperiment2,TrionExperiment3,TrionExperiment4,TrionExperiment5,TrionExperiment6,TMDCDEmler} have 
demonstrated that in the presence of carriers the prominent excitonic ($\hbox{X}$) features 
in optical absorption split into two separate peaks.
This property is closely related to the carrier-induced
splitting of up to $\sim 2~ \hbox{meV}$ observed 
previously in conventional $\hbox{GaAs}$ and $\hbox{CdTe}$~\cite{QW1,QW2,QW3,QW4,QW5} quantum wells, 
but can be ten or more times larger, allowing it to be resolved at higher temperatures. 
The appearance of an additional peak is usually attributed to the presence of trions ($\hbox{T}$), charged fermionic 
quasiparticles formed by binding two electrons to one hole or two holes to one electron. 
The splitting energy often coincides approximately with theoretical~\cite{TrionExperiment4,TrionTheory1,TrionTheory2,TrionTheory3,TrionTheory4}
trion binding energies, $\epsilon_\mathrm{T}$, supporting this interpretation.  
A full theory of trion absorption that could establish this scenario more definitively wouldhowever need to account for higher energy three-particle bound states and for the matrix elements of optical 
transitions between trion and single-particle states, and is absent at present.  

\begin{figure}[t]
\begin{center}
	\label{Fig1}
	\vspace{-0.5cm}
	\includegraphics[trim=0.4cm 4.6cm 2cm 1.5cm, clip, width=1.0\columnwidth]{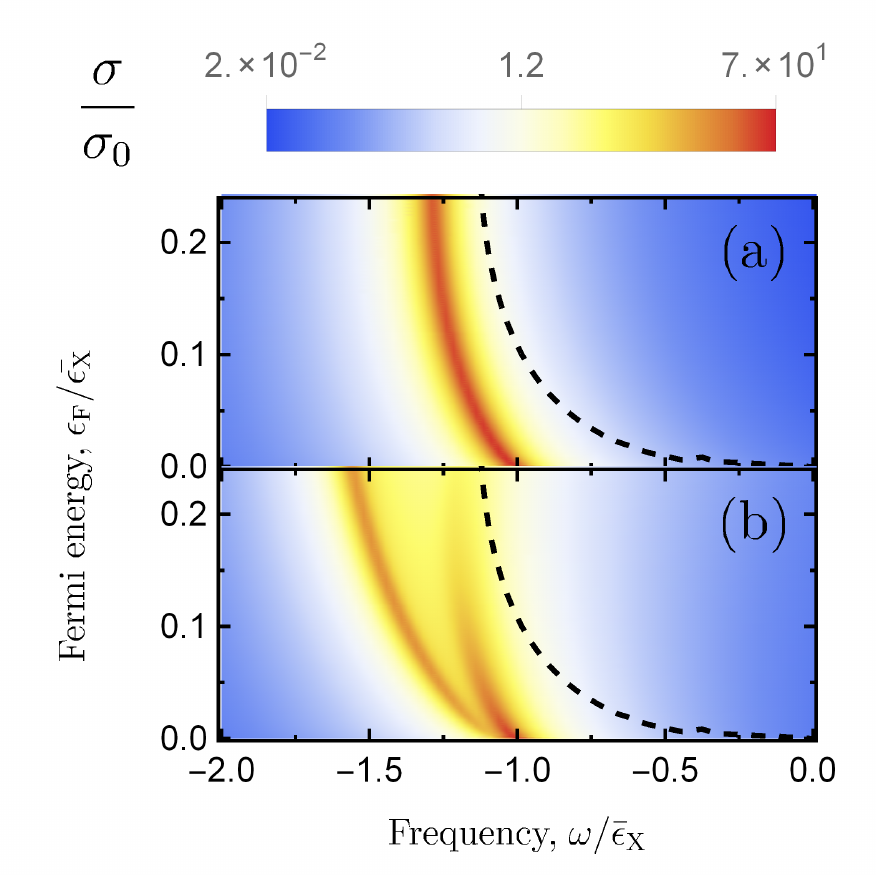}
	\vspace{-0.5cm}
	\caption{Optical conductivity $\sigma(\omega)/\sigma_0$ where $\sigma_0=e^2/h$ is the 
	quantum unit of conductance.   
$\hbox{(a)}$ theoretical conductivity when Fermi-sea dressing of excitons is neglected  
and $\hbox{(b)}$ full conductivity including interactions between excitons and a fluctuating Fermi sea.
We refer to the two peaks in (b), often interpreted as trion and exciton peaks,
as the attractive and repulsive exciton-polaron branches.
Energies are measured from the bare semiconductor band gap and measured in units of the 
exciton binding energy.  The dashed lines show the bare
interband absorption threshold renormalized by interactions.
%exciton-polaron modes.  The dressed exciton modes have 
}
\end{center}\vspace{-0.5cm}

\end{figure}

There is in fact substantial doubt~\cite{Combescot1,Combescot2} that 
the absorption spectrum can be adequately interpreted in terms of 
three-body physics. The reason is that a three-particle description is valid only at low-doping $\epsilon_\mathrm{F}\ll\epsilon_\mathrm{T}$, where $\epsilon_\mathrm{F}$ is the Fermi level of the excess charge carriers. The additional peak is clearly observed experimentally only at an intermediate level with  $\epsilon_\mathrm{F}\sim\epsilon_\mathrm{T}$,  but still small compared to the excition binding energy $\epsilon_\mathrm{X}$. It has been argued on physical grounds that a picture of excitons interacting with excitations of Fermi sea is more appropriate~\cite{Wouters1,Wouters2,TMDCDEmler,Suris1, Suris2}. Recently it has been explained by Sidler {\em et al.}~\cite{TMDCDEmler} that the main effect of these interactions is dressing of excitons to exciton-polarons. In the present work we provide a detailed microscopic theory of exciton-polarons, and demonstrate that its predictions are in good agreement with experiment.

Our main results for the dependence of optical conductivity on frequency and carrier density 
are summarized in Fig.~1.  The main absorption features lie well below the non-interacting-particle
absorption threshold over a wide carrier-density range.
The relevant low-energy degrees of freedom are therefore the exciton's center 
of mass, and excitations of the Fermi sea.  Because of their mutual interactions,
the excitonic state splits into attractive and repulsive exciton-polaron branches,
which are many-body generalizations of trion bound and unbound states respectively.
The splitting between peaks is linear in carrier density and the excitonic peak
broadens and smoothly disappears as carrier density increases, in good agreement with experiment.

The rest of the paper is organized as follows. In Sec.~II the minimal model sufficient to describe optical properties of TMDC is introduced. In Sec.~III we introduce excitons and calculate their contribution to optical conductivity.  In Sec.~IV the dressing of excitons to exciton-polaron is presented. Sec.~V presents doping dependence of optical conductivity. We summarize in Sec.~VI.

\noindent
\section{II. 2D semiconductor model}---
The optical properties of two-dimensional TMDCs can be described using a parabolic band model 
with electron and hole carriers in two valleys $\alpha=\pm 1$~\cite{CC2}.
The single valley Hamiltonian is given by  
\begin{equation*}
H=\sum_{\vec{p}\gamma}\epsilon_{\vec{p}}^\gamma a_{\vec{p}\gamma}^+ a_{\vec{p}\gamma} + \frac{1}{2}  \sum_{\gamma \gamma'}\sum_{\vec{p}\vec{p}'\vec{q}} V^0_\vec{q} a_{\vec{p}+\vec{q}, \gamma}^+ a_{\vec{p}'-\vec{q},\gamma'}^+a_{\vec{p}' \gamma'} a_{\vec{p}\gamma},
\end{equation*}
%\begin{equation*}
%\begin{split}
%H=\sum_{\vec{p}} \left[\epsilon_{\vec{p}}^\mathrm{c} a_{\vec{p}}^+ a_{\vec{p}} + \epsilon_{\vec{p}}^\mathrm{v} b_{\vec{p}}^+ b_{\vec{p}} \right]+ %\sum_{\vec{p}\vec{p}'\vec{q}} V^0_\vec{q} a_{\vec{p}+\vec{q}}^+ b_{\vec{p}'-\vec{q}}^+b_{\vec{p}'} a_\vec{p}\\+
%\frac{1}{2}\sum_{\vec{p}\vec{p}'\vec{q}} V^0_\vec{q} a_{\vec{p}+\vec{q}}^+ a_{\vec{p}'-\vec{q}}^+a_{\vec{p}'} a_\vec{p} + %\frac{1}{2}\sum_{\vec{p}\vec{p}'\vec{q}} V^0_\vec{q} b_{\vec{p}+\vec{q}}^+ b_{\vec{p}'-\vec{q}}^+b_{\vec{p}'} b_\vec{p}.
%\end{split}  
%\end{equation*}
where $\gamma=\mathrm{c},\mathrm{v}$ denotes electrons from conduction and valence bands with dispersion laws $\epsilon_\vec{p}^\mathrm{c}=\vec{p}^2/2m -\epsilon_\mathrm{F}$ and 
$\epsilon_\vec{p}^\mathrm{v}=-\vec{p}^2/2m-\epsilon_\mathrm{g}-\epsilon_\mathrm{F}$, $\epsilon_\mathrm{g}$ is the energy gap, 
$V_\vec{q}^0=2\pi e^2/\kappa q$ is the bare Coulomb interactions, and $\kappa$ 
is the dielectric constant of TMDC material~\cite{CC6}. 
We describe the light matter interaction using a position independent vector potential $\vec{A}$:
\begin{equation}
H_\mathrm{EM}=-\frac{e v}{c}\sum_{\vec{p}\alpha} \vec{A} \cdot\left[ \vec{e_\alpha}a_{\vec{p}\mathrm{c}\alpha}^+a_{\vec{p}\mathrm{v}\alpha} e^{- \bm i (\omega + \epsilon_\mathrm{g}) t} + \mathrm{h}.\mathrm{c}. \right]. 
\end{equation}
Here $v=(\epsilon_\mathrm{g}/2m)^{1/2}$ is the matrix element of the velocity operator 
between conduction and valence bands~\cite{CC4}, $\omega$ is the photon energy measured from the semiconductor band gap,
%gap $\epsilon_\mathrm{g}$, as we do for kinetic energy of electrons. 
and the valley-dependent vector $\vec{e}_\alpha=\vec{e}_x+\alpha \;\bm{i}\vec{e}_y$ 
encodes the spin-valley locking property of two-dimensional semiconductors that 
enables valley-selection using circularly polarized light.
%It should be noted that spin-valley locking is of importance here and our results are well applicable for conventional two-dimensional systems in semiconductor quantum wells.    

\noindent
\section{III. Bare excitonic states}
The formulation of our theory of optical conductivity requires that we first 
consider the artificial limit in which
Fermi sea fluctuations are suppressed.  In order to establish needed notation 
we first briefly describe that limit, while the detailed derivations are presented in Appendix A for completeness. The optical conductivity can be expressed
as a sum over total momentum $\vec{q}=0$ excitonic (and scattering electron-hole) states
% are labeled by principle $n=0,1,..$ and 
%orbital $m=-n,..,n$ quantum numbers, and 
which satisfy relative-motion Schrodinger equations that have the following 
momentum-space form:  
\begin{equation}
\label{ExcitonEigenvalue}
\left[\frac{\vec{p}^2}{2 \mu_\mathrm{X}}+\Sigma_\mathrm{g}\right] C_\vec{p}-\sum_{ \vec{p}'} B_\vec{p} V_{\vec{p}-\vec{p}'} B_{\vec{p}'}C_{\vec{p}'}=\epsilon_\mathrm{X} C_\vec{p}. 
\end{equation}
Here $C_\vec{p}$ and $\epsilon_\mathrm{X}$ are the exciton momentum space wave functions and energies, 
$\mu_\mathrm{X}=m/2$ is the reduced mass,
and $B_\vec{p}=[1-n_{\mathrm{F}}(\epsilon^\mathrm{c}_\vec{p})]^{1/2}$ is a Pauli blocking factor that 
excludes filled electronic states from the space available for exciton formation.
For screening we use the static random phase approximation (RPA), $
V_{\vec{p}}=2 \pi e^2/\kappa[p+p_\mathrm{sc}(p)]$,
with screening momentum given by 
$p_\mathrm{sc}(p)=-2\pi e^2 \Pi(p)/\kappa$ with the static polarization operator of two-dimensional electron gas $\Pi(p)=- m/\pi \hbar^2 \times \{1-\Theta(p-2 p_\mathrm{F})[1-(2p_\mathrm{F}/p)^2]^{1/2}\}$.~\cite{PolarizationOperator}
In Eq.~(\ref{ExcitonEigenvalue}) $\Sigma_\mathrm{g}$ accounts for gap renormalization by carriers due to 
screening and phase-filling effects:
\begin{equation}
\Sigma_\mathrm{g}=-\sum_{\vec{p}} V_\vec{p} n_\mathrm{F}(\epsilon_\vec{p}^\mathrm{c})- \sum_{\vec{p}}(V_\vec{p}^0-V_\vec{p}) n_\mathrm{F}(\epsilon_\vec{p}^\mathrm{v}).
\end{equation}

When the gap renormalization is included, 
the single-particle absorption threshold $2 \epsilon_\mathrm{F}+\Sigma_\mathrm{g}$ 
is red shifted by electron-electron interactions.

\begin{figure}[t]
	\label{Fig2}
	\vspace{-2 pt}
	\includegraphics[width=7.7 cm]{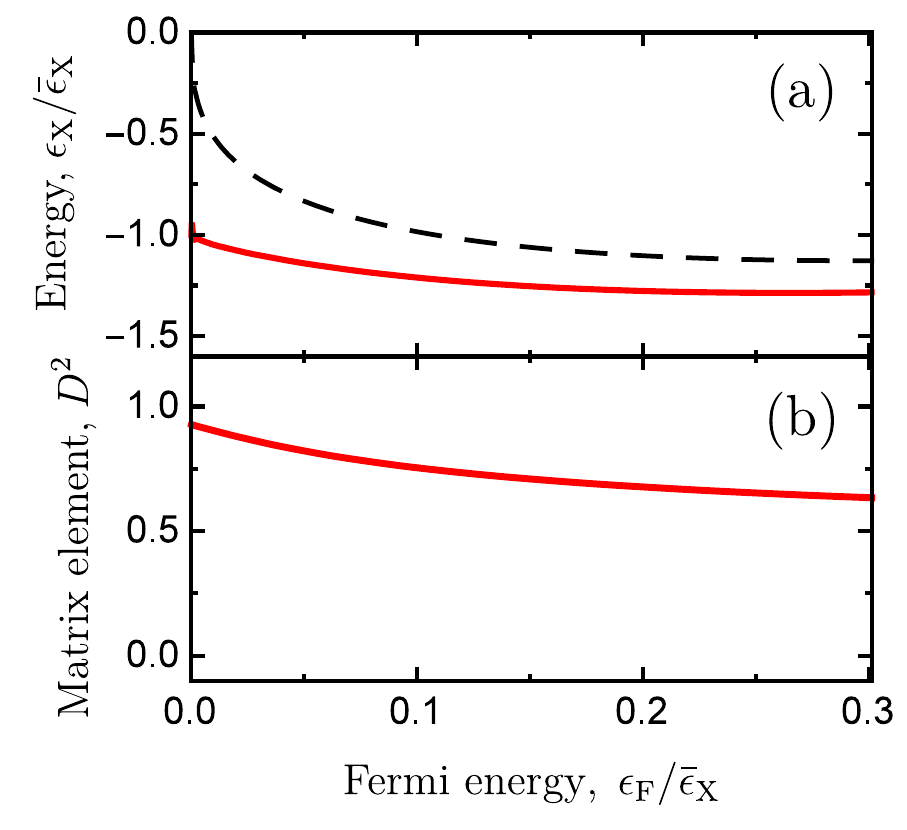}
	\vspace{-2 pt}
	\caption{Dependence on carrier Fermi energy $\epsilon_\mathrm{F}$
	of $\hbox{(a)}$ the excitonic ground state energy $\epsilon_\mathrm{X}$ and $\hbox{(b)}$ its
     squared optical matrix element $D^2$.  The ground state 
	energy approaches the renormalized interband absorption threshold  
	$2 \epsilon_\mathrm{F}+\Sigma_\mathrm{g}$ (dashed line) in the high carrier density limit.}
\end{figure}

When carriers are absent the eigenvalue equation (\ref{ExcitonEigenvalue}) 
maps to the two-dimensional hydrogenic Schrodinger equation 
which has an analytic solution with bound state 
energies $\epsilon_\mathrm{X}^{nm_\mathrm{z}}=-\bar{\epsilon}_\mathrm{X}/(2n+1)^2$, where $n,m_\mathrm{z}$ are main and orbital quantum numbers. Here $\bar{\epsilon}_\mathrm{X}=m e^4/\kappa^2 \hbar^2 = 4 Ry^{*}$ is the ground state
binding energy and $Ry^*$ is the excitonic Rydberg energy. 
%Wave function stretch in  momentum space, or the inverse exciton radius, is given by $\bar{p}_\mathrm{X}=me^2/\kappa \hbar$. 
When carriers are present the eigenvalue problem (\ref{ExcitonEigenvalue}) 
must be solved numerically. The rotational symmetry allows to label bound states in the same way, and their momentum dependence can be factorized as follows $C_\vec{p}^{nm_\mathrm{z}}=C^{nm_\mathrm{z}}(p) \mathrm{exp}[\bm{i} m_\mathrm{z} \phi_\vec{p}]/\sqrt{2\pi}$. The dependence of the ground state binding energy,
$\epsilon_\mathrm{X}^{00}\equiv\epsilon_\mathrm{X}$, on carrier Fermi energy $\epsilon_\mathrm{F}$ 
that results from these approximations is illustrated in Fig.~2-a. 
The binding energy smoothly decreases with doping and the excitonic state asymptotically approaches the absorption threshold $2 \epsilon_\mathrm{F}+\Sigma_\mathrm{g}$. It does not merge with the threshold since in two space 
dimensions bound states are formed for arbitrarily weak attractive interactions. 
Higher energy excitonic bound states play little role when carriers are present;
we find that the last excited bound state $\epsilon_\mathrm{X}^{10}$ already merges with the
continuum at $\epsilon_\mathrm{F}/\bar{\epsilon}_\mathrm{X}\approx 0.02$.
Because we are interested in the sharp bound sate absorption features, 
we do not focus on scattering states,which govern the absorption above threshold. 
% omit scattering states %We do not focus on the scattering state contribution to the absorption spectrum because it is difficult to isolate experimentally. 
 
When fluctuations of the Fermi sea are neglected the optical conductivity 
\begin{equation}\label{OpticalConductivity}
\sigma(\omega)=2 \sigma_0 \sum_{nm_\mathrm{z}} \left|D^{nm_\mathrm{z}}\right|^2 \; \bar{\epsilon}_\mathrm{X} A^{nm_\mathrm{z}}_\mathrm{X}(\omega,0),
\end{equation}
where $\sigma_0=e^2/h$ is the conductivity quantum, $M_\mathrm{X}=2m$ is the total exciton mass,
$A^{nm_\mathrm{z}}_\mathrm{X}(\omega,\vec{q})=-2 \mathrm{Im}[G_{\mathrm{X}}^{nm_\mathrm{z}}(\omega,\vec{q})]=-2 \mathrm{Im}[(\omega_+ - \epsilon^{nm_\mathrm{z}}_\mathrm{X}-\vec{q}^2/2 M_\mathrm{X})^{-1}]$ is the spectral function of excitons 
in state $n,m_\mathrm{z}$, and $\omega_+ = \omega + \bm{i} \gamma$ includes a phenomenologically introduced finite-lifetime energy 
uncertainty $\gamma$.  
Here $D$ is the dimensionless optical coupling matrix element
\begin{equation}\label{MatrixElement}
D=\sqrt{\frac{\pi}{2 \bar{p}_\mathrm{X}^2}} \sum_\vec{p}  B_\vec{p} C_\vec{p},
\end{equation}
which is non-zero only for states with $m_\mathrm{z}=0$, since $B_\vec{p}$ does depend only on absolute value of $\vec{p}$.   
The ground state matrix element $D$ decreases slowly with carrier density, as illustrated Fig.~2-b,
and the corresponding optical conductivity $\sigma$ is plotted in Fig.~1-a. 
The excitonic peak slowly weakens and shifts toward the continuum absorption edge as 
the carrier density increases.       

\begin{figure}[t]
  	\label{Fig3}
  	\vspace{-2 pt}
  	\includegraphics[width=7.7 cm]{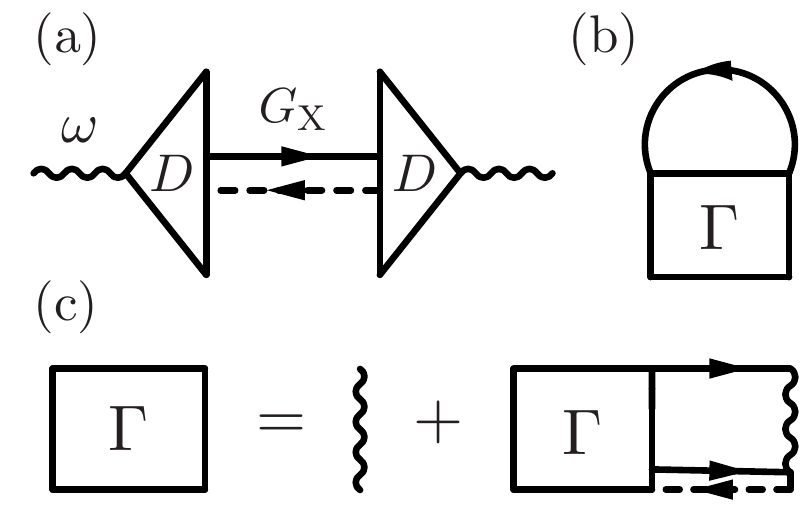}
  	\vspace{-2 pt}
  	\caption{$\hbox{(a)}$ Excitonic contribution to the optical conductivity.  The triangle vertexes correspond to the
	optical matrix elements $D$, defined in Eq.~\ref{MatrixElement}. The paired solid and dashed lines represent excitons, bound states of 
	conduction band electrons and valence band holes described by summation of all scattering ladder diagrams. $\hbox{(b)}$ 
	Exciton self-energy due to interactions $\Gamma$ with Fermi sea fluctuations. $\hbox{(c)}$ Bethe-Salpeter equation for
	the exciton/Fermi-sea interaction $\Gamma$-vertex.   The value of these diagrams depends on the total and relative motion momenta 
	$\vec{q}$ and $\vec{p}$.  For a given $\vec{q}$ and $\vec{p}$ the exciton momentum is $2\vec{q}/3-\vec{p}$ and the electron 
	momentum is $\vec{q}/3+\vec{p}$.}
\end{figure}  

\noindent
\section{IV. Exciton-polarons} The optical conductivity has previously been studied extensively in the absence of carriers,
when Eq.~(\ref{OpticalConductivity}) applies, and in the high carrier density limit when 
$\epsilon_\mathrm{F}\sim \bar{\epsilon}_\mathrm{X}$ and the theory of Fermi edge singularities~\cite{Mahan1,Mahan2,OpticsReview}
applies.  In this Letter we focus on the intermediate regime in which $\epsilon_\mathrm{F}\sim\epsilon_\mathrm{T}\ll \bar{\epsilon}_\mathrm{X}$
and the excitonic peak is still far from the absorption edge.  In this regime the low-energy degrees-of-freedom are those with 
an energy below $\bar{\epsilon}_\mathrm{X}$, namely the excitonic center of mass and carrier Fermi-sea fluctuations.   
The interactions between these two types of degrees of freedom lead to dressed excitons that we 
refer to as exciton-polarons. 

Because of the valley degeneracy,
two Fermi seas disturb the 
excitons. When the excitons and carrier Fermi seas are associated with the same valley they 
have short-range repulsive exchange interactions which limit correlations. In the low density regime $\epsilon_\mathrm{F}\ll \epsilon_\mathrm{T}$ exchange interactions do not favor the formation of trion states, except for the case of strong imbalance between masses of electron and hole not realized in TMDC~\cite{Suris3}. In the considered density range $\epsilon_\mathrm{F}\sim \epsilon_\mathrm{T}$ the exchange interactions are even more important, so we assume that excitons are dressed by the Fermi sea only from the different valley. 
The condition $\epsilon_\mathrm{F}\ll \bar{\epsilon}_\mathrm{X}$ implies 
that the electrons are too dilute to unbind the excitons, polarizing them instead to induce attractive interactions.
Below we approximate these interactions by short-range ones with momentum independent Fourier transform $U$. We estimate it and $\epsilon_T/\bar{\epsilon}_\mathrm{X}$ in Appendix B and show that this approximation is reasonable. Nevertheless it is instructive to treat $U$ as an independent parameter in our model.

Our approximation for the full optical conductivity is summarized in Fig.3.  
Eq.~(\ref{OpticalConductivity}), which is exact in the absence of carriers, is summarized schematically 
in Fig.~3a.  When Fermi sea fluctuations are included the exciton propagator in Eq.~(\ref{OpticalConductivity}) is dressed by the self-energy in Fig.~3b which accounts for the attractive interaction between
excitons and Fermi sea electrons by summing the ladder diagrams.  
A similar approximation~\cite{FermiPolaronReview, FermiPolaronDemler} has recently been used to 
describe dilute minority spins in a fermionic cold atom 
majority spin gas.
The two-particle scattering function $\Gamma^\mathrm{R}(\omega,\vec{q})$ in Fig.~3c satisfies 
a Bethe-Salpeter equation, $\Gamma^\mathrm{R}=U+UK^\mathrm{R}\Gamma^\mathrm{R}$,
which simplifies to an algebraic equation when the momentum and frequency dependence of $U$ is neglected.  
In this approximation, the kernel 
\begin{equation}
\label{KKernel}
K^\mathrm{R}(\omega,\vec{q})=\sum_\vec{p}\frac{1-n_\mathrm{F}(\epsilon^\mathrm{c}_{\vec{p}+\vec{q}/3})}{ \omega_+-\epsilon_\mathrm{X}-\frac{\vec{q}^2}{2 M_\mathrm{T}}-\frac{p^2}{2 \mu_\mathrm{T}}+\epsilon_\mathrm{F}}.
\end{equation}
depends only on the total incoming momentum $\vec{q}$ and frequency $\omega$. 
In Eq.~(\ref{KKernel}) $M_\mathrm{T}=3m$ and $\mu_\mathrm{T}=2m/3$ are the
total and reduced masses of the exciton-electron system.
Generalizing the calculations in Refs.~\cite{FermiPolaronDemler,FermiPolaron1,FermiPolaron2}
to the case of unequal mass ($m$ and $2m$) particles, we find that  
\begin{equation}
\label{GammaVertex}
\Gamma^\mathrm{R}(\omega,\vec{q})=\frac{2\pi \hbar^2}{\mu_\mathrm{T}}\frac{1}{\log\left[\frac{\epsilon_\mathrm{T}}{\Omega}\right]+\bm{i} \pi},
\end{equation}
where $\epsilon_\mathrm{T}=p_\mathrm{\Lambda}^2/2\mu_\mathrm{T} \times \exp[-2 \pi \hbar^2/\mu_\mathrm{T} U]$ is
the trion binding energy in the absence of carriers and $p_\mathrm{\Lambda}$ is a momentum-space  ultraviolet cutoff. 
Using this equation we are able to express $\Gamma^\mathrm{R}$ in terms of the trion binding energy alone, eliminating $U$ and ultraviolet cutoff $p_\mathrm{\Lambda}$ from the theory.
In Eq.~(\ref{GammaVertex}) the energy $\Omega$ is given by
\begin{equation*}
\begin{split}
\Omega=\frac{1}{2} \Biggl\{\omega_+-\epsilon_\mathrm{X} -\frac{\vec{q}^2}{4 M_\mathrm{T}}-\frac{p_\mathrm{F}^2}{4m} + s\times\quad \quad \quad \quad \quad   \\  \sqrt{\left[ \omega_+-\epsilon_\mathrm{X}-\frac{(p_\mathrm{F}+q)^2}{4m} \right]\left[\omega_+-\epsilon_\mathrm{X}-\frac{(p_\mathrm{F}-q)^2}{4m} \right]} \Biggr\},
\end{split}
\end{equation*}
where $s=\mathrm{sign}(\omega-\epsilon_\mathrm{X}-p_\mathrm{F}^2/4m-q^2/4m)$. It is instructive to introduce the molecular spectral function for excitons and electrons as $A_\mathrm{\Gamma}(\omega,\vec{q})=-2 \Im[\Gamma^\mathrm{R}(\omega,\vec{q})]$. It is presented at different doping levels in Fig.~4. The spectral function is nonzero within the continuum of excited exciton-electron states and has a single separate peak along the dispersion curve $\omega_\vec{q}$, which corresponds to their bound state and is given by
\begin{equation}
\label{WqS}
\omega_\vec{q}=\epsilon_\mathrm{X}-\frac{\left(\epsilon_\mathrm{T}-\frac{\vec{q}^2}{2M_\mathrm{T}}\right)\left(\epsilon_\mathrm{T}-\frac{p_\mathrm{F}^2}{4m}+\frac{\vec{q}^2}{4M_\mathrm{T}}\right)}{\epsilon_\mathrm{T}+\frac{\vec{q}^2}{4M_\mathrm{T}}}.
\end{equation}

\begin{figure}[t]
	\label{Fig4}
	\vspace{-2 pt}
	\includegraphics[trim=1.2cm 6.2cm 1.2cm 0cm,width=5.7 cm]{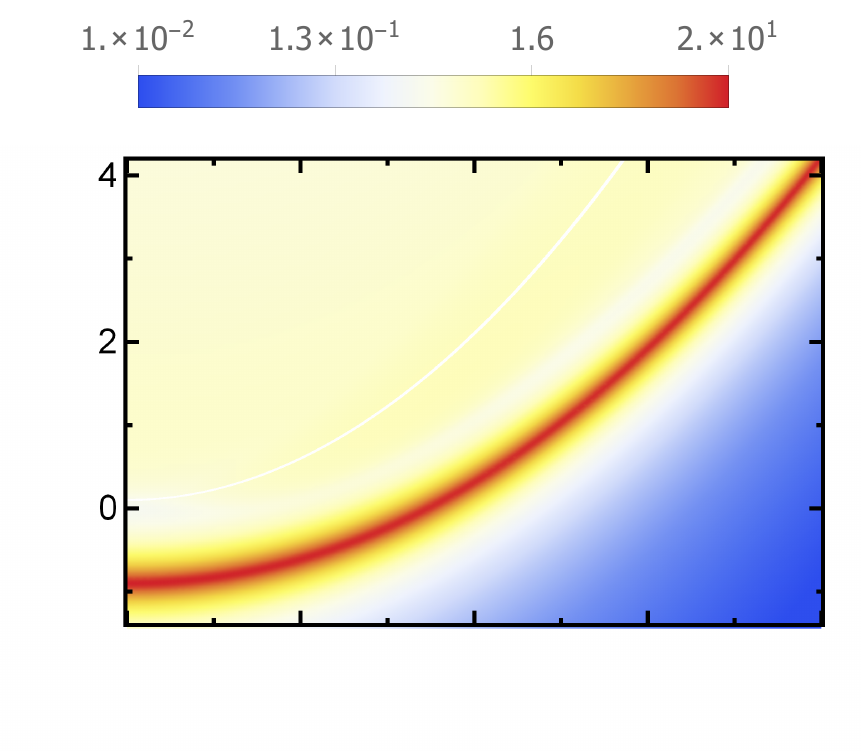}
	\includegraphics[trim=1.2cm 0cm 1.2cm 0cm,width=7.6 cm]{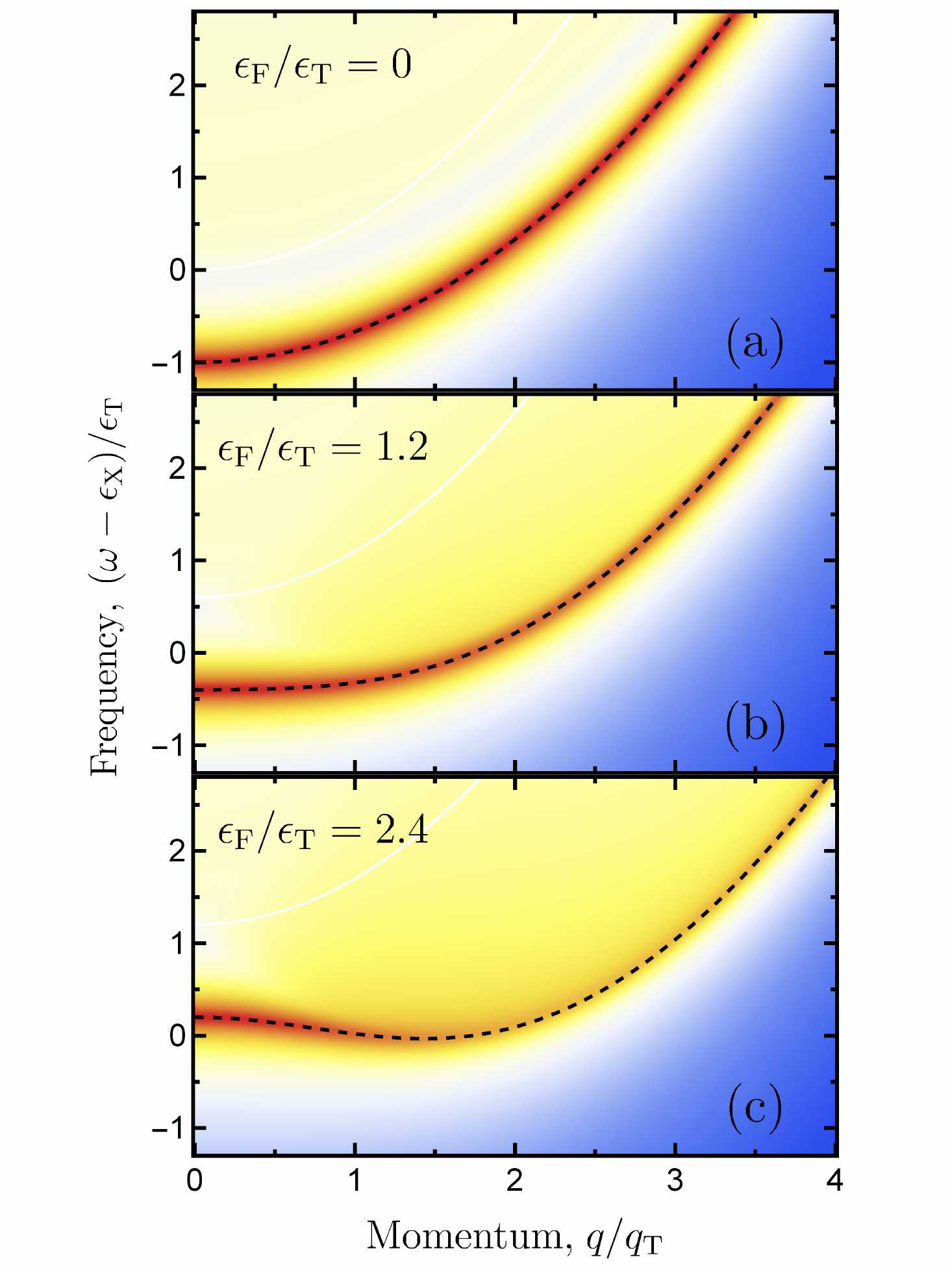}
	\vspace{-2 pt}
	\caption{The spectral function  $A_\mathrm{\Gamma}(\omega,\vec{q})=-2\Im[\Gamma^\mathrm{R}(\omega,\vec{q})]$ for the many-body vertex function $\Gamma^\mathrm{R}(\omega,\vec{q})$. The dashed line follows $\omega_\vec{q}$, given by  (\ref{WqS}) and corresponding to the bound state of exciton with Fermi sea of electrons.  The behavior evolves from the two-particle one at $\epsilon_\mathrm{F}\ll\epsilon_\mathrm{T}$ to the polaronic one $\epsilon_\mathrm{F}\sim\epsilon_\mathrm{T}$, where the dispersion $\omega_\vec{q}$ achieves minimum at finite momentum $q_*$ and can be expanded in its vicinity according to (\ref{WqSMin1}) and (\ref{WqSMin2}).} 
\end{figure} 
At $\epsilon_\mathrm{F}\ll \epsilon_\mathrm{T}$ the dispersion law simplifies to $\omega_\vec{q}=\epsilon_\mathrm{X}-\epsilon_\mathrm{T}+\vec{q}/2 M_\mathrm{T}$ and represents the two-particle behavior. Moreover, many-body $\Gamma$-vertex reduces to two-particle $T$-matrix for scattering of electron and exciton. In the polaronic regime at $\epsilon_\mathrm{F}>\epsilon_0$, with $\epsilon_0=4 \epsilon_\mathrm{T}/3$, the dispersion law $\omega_\vec{q}$ achieves minimum at finite momentum $q_*$ and can be expanded in its vicinity as follows
\begin{equation}
\label{WqSMin1}
\omega_\vec{q}\approx\epsilon_\mathrm{X}-\epsilon_*+\frac{(q-q_*)^2}{2 m_*},
\end{equation}
where the binding energy $\epsilon_*$, effective mass $m_*$ and the finite momentum $q_*$ of the exciton-polaron state are given by
\begin{equation}
\label{WqSMin2}
\epsilon_*=\frac{4 \epsilon_\mathrm{T}}{3} \left(\frac{p_\mathrm{F}}{p_0}-\frac{3}{2}\right)^2,\;  m_*=\frac{3 m}{4} \left(\frac{p_\mathrm{F}}{p_0}-1\right)^{-1},
\end{equation}
\vspace{-0.4cm}
\begin{equation} \mathrm{q}_*=\sqrt{6} q_\mathrm{T} \left(\frac{p_\mathrm{F}}{p_0}-1\right)^{\frac{1}{2}}.
\end{equation}
Here we introduced the momentum $p_0= \sqrt{2m \epsilon_0}=2 q_\mathrm{T}/\sqrt{3}$. Note that the binding energy $\epsilon_*$ is always positive, making the formation of the exciton-electron bound state energy favorable, and the mass $m_*$ diverges at $\epsilon_\mathrm{F}= \epsilon_0$. 
The continuum of excited states also evolves from the two-particle behavior, where the bound state peak and the boundary of continuum are well-separated, to the polaronic behavior, where the continuum and the dispersion $\omega_\vec{q}$ of the exciton-electron bound state almost merge with each other. It should be noted that the spectral function for excitons and electrons $A_\mathrm{\Gamma}(\omega,\vec{q})$ contains a lot of information about the polaronic physics~\cite{FermiPolaronDemler,FermiPolaronReview}. Nevertheless, it is not probed directly in the absorption experiments, but the spectral function of excitons at zero momentum $A_\mathrm{X}(\omega,0)$, which is connected with the $\Gamma$-vertex in the nontrivial way.

Finally, to evaluate the optical absorption using Eq.~(\ref{OpticalConductivity}) we need to calculate the excitonic 
spectral function at momentum
$\vec{q}=0$:   
$A_\mathrm{X}(\omega,0)=-2 \mathrm{Im}\left[\left\{\omega_+-\epsilon_\mathrm{X}-\Sigma^\mathrm{R}_\mathrm{X}(\omega_+,0)\right\}^{-1}\right],$ where in the approximation of Fig.~3c 
\begin{equation}
\Sigma_\mathrm{X}^\mathrm{R}(\omega,0)=\sum_{\vec{p}} \Gamma^\mathrm{R}(\omega+\epsilon_{\vec{p}'}^\mathrm{c},\vec{p})n_\mathrm{F}(\epsilon_{\vec{p}}^\mathrm{c}).
\label{SelfEnergy}
\end{equation}
This self-energy is responsible for a peak in the exciton spectral weight close to
the trion energy whose weight vanishes in the limit of zero carrier density.   

\section{V. Results} 
Our theory expresses the conductivity in terms of five energy scales, the disorder scale $\gamma$, 
the exciton binding energy $\bar{\epsilon}_{\mathrm{X}}$, the trion binding energy $\epsilon_\mathrm{T}$, 
the Fermi energy of electrons $\epsilon_\mathrm{F}$ and the photon energy $\omega$. 
For the results presented below we fix $\gamma/\bar{\epsilon}_\mathrm{X}=0.03$, and in agreement with experiment choose  
$\epsilon_\mathrm{T}/\bar{\epsilon}_\mathrm{X}\approx 0.07$.
We also presents these plots in real units in Appendix C for completeness. With these ratios fixed we
calculate the dependence of the theoretical conductivity on $\epsilon_F$ and $\omega$ which
we have illustrated in Fig.1-b. Its sections are presented in Fig.~5   
The self-energy, (\ref{SelfEnergy}), mixes excitons and Fermi sea excitations and leads to 
two peaks in optical absorption   
that can be associated with attractive and repulsive polaronic branches, which are many-body generalizations of trion bound and unbound states. In the low-carrier density limit, the two absorption peaks
correspond precisely to the excitation of trions and excitons at energies $\epsilon_\mathrm{T}^*$ and $\epsilon_\mathrm{X}^*$
respectively~\cite{CC5}. The $^*$ accents here emphasize that the binding energies are renormalized in a non-trivial way at finite Fermi energy $\epsilon_\mathrm{F}$.  To preserve the conventional terminology we refer to these peaks as to exciton and trion ones.

\begin{figure}[t]
	\label{Fig5}
	\includegraphics[width=8.5 cm]{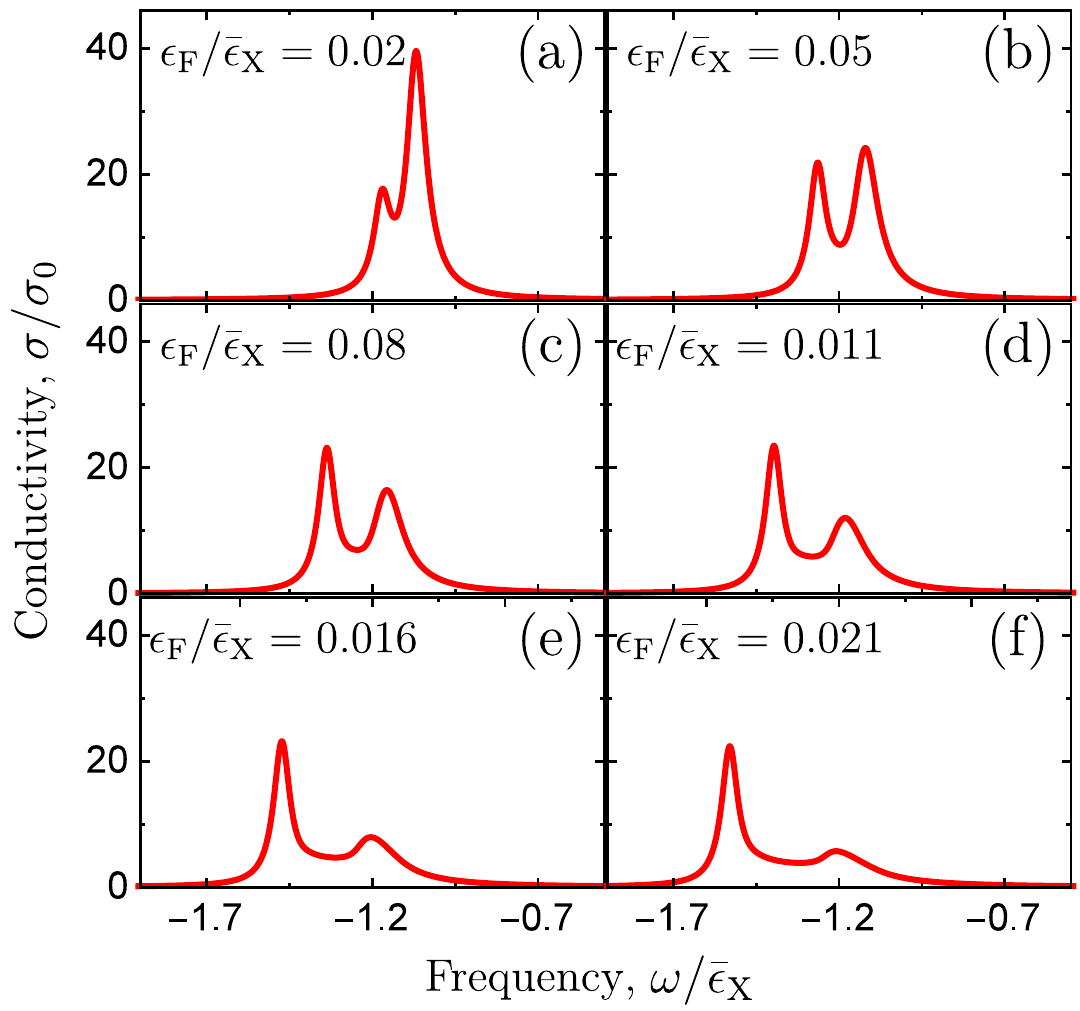}
	\vspace{-2 pt}
	\caption{$\hbox{(a)}$-$\hbox{(f)}$ Frequency dependence of the optical conductivity $\sigma(\omega)$ for different values of the Fermi energy of electrons $\epsilon_\mathrm{F}$. The trion binding energy is equal to $\epsilon_\mathrm{T}/\epsilon_\mathrm{X}=0.07$. Two peaks represent attractive and repulsive exciton-polaron branches. }
\end{figure}

\begin{figure}[t]
	\label{Fig6}
	\vspace{-2 pt}
	\includegraphics[width=7.7 cm]{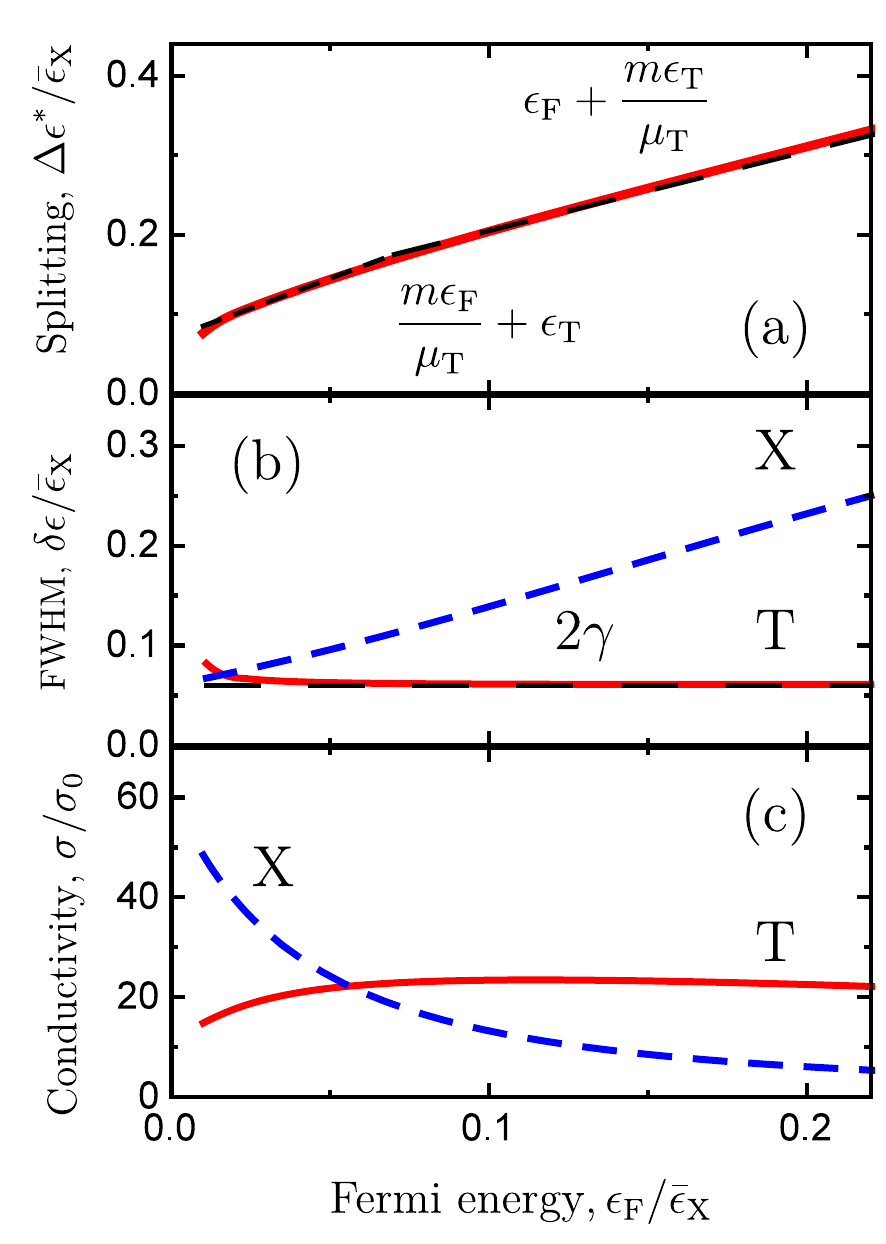}
	\vspace{-2 pt}
	\caption{$\hbox{(a)}$ Dependence of the energy splitting $\Delta \epsilon^*=\epsilon_\mathrm{X}^*-\epsilon_\mathrm{T}^*$ between exciton $\hbox{X}$ and trion $\hbox{T}$ absorption features on carrier Fermi energy.
	 Dependence of absorption feature $\hbox{(b)}$  peaks $\sigma(\epsilon_\mathrm{X(T)}^*)/\sigma_0$  and
	$\hbox{(c)}$ widths $\delta \epsilon_\mathrm{X(T)}/\bar{\epsilon}_\mathrm{X}$  on carrier Fermi energy $\epsilon_\mathrm{F}$. 
	The splitting interpolates between two linear behaviors $\Delta\epsilon^*= \epsilon_\mathrm{T}+m\epsilon_\mathrm{F}/\mu_\mathrm{T}$ at $\epsilon_\mathrm{F}\lesssim \epsilon_\mathrm{T}$, and $\Delta\epsilon^*= m\epsilon_\mathrm{T}/\mu_\mathrm{T} +\epsilon_\mathrm{F}$ at $\epsilon_\mathrm{F}\gtrsim \epsilon_\mathrm{T}$.}
\end{figure}

Before discussion of the doping dependence of the absorption, it is constructive to consider low carrier-density limit $\epsilon_\mathrm{F} \ll \epsilon_\mathrm{T}$ . In that limit the exciton-electron problem reduces to a two-particle one and the 
excitonic self-energy and spectral function can be calculated analytically. Details of derivation are presented in Appendix D. We find that to leading order in $\epsilon_{F}/\epsilon_{T}$, 
$A_\mathrm{X}(\omega,0)\approx 2\pi Z_\mathrm{T} \delta(\omega-\epsilon_T^*)+ 2\pi Z_\mathrm{X}\delta(\omega-\epsilon_X^*)$, where $\epsilon_\mathrm{T}^*=\epsilon_\mathrm{X}-\epsilon_\mathrm{T}-m \epsilon_\mathrm{F}/\mu_\mathrm{T}$ and $\epsilon_\mathrm{X}^*=\epsilon_\mathrm{X}$ are positions of peaks. $Z_\mathrm{T}= m \epsilon_\mathrm{F}/\mu_\mathrm{T}\epsilon_\mathrm{T}$ and $Z_\mathrm{X}=1-Z_\mathrm{T}$ are their spectral weights. The splitting between peaks goes linearly $\Delta \epsilon^*=\epsilon_\mathrm{T}+m \epsilon_\mathrm{F}/\mu_\mathrm{T}$ with Fermi energy of electrons, while its value at zero doping equal to the trion binding energy $\epsilon_{T}$. The trion peak spectral weight $Z_\mathrm{T}$ vanishes in the absence of doping and, the most importantly, is much smaller then one of exciton as long as $\epsilon_\mathrm{F}\ll \epsilon_\mathrm{T}$. Although our model of a trion as a bound state of an exciton and an electron is simplified, this relation between spectral weight can be rigorously established. We conclude that the competition between peaks can not be attributed to three-particle physics.

The dependence of splitting between exciton and trion peaks on the Fermi energy $\epsilon_\mathrm{F}$ of electrons is presented in Fig.~6-a. We see there that for $\epsilon_\mathrm{F}\lesssim \epsilon_\mathrm{T}$, the splitting goes linearly with the Fermi energy as $\Delta\epsilon^*= \epsilon_\mathrm{T}+m\epsilon_\mathrm{F}/\mu_\mathrm{T}$, which is consistent with our analytical results. 
It is notable that at $\epsilon_\mathrm{F}\gtrsim \epsilon_\mathrm{T}$ the dependence evolves to another linear behavior with a 
different slope $\Delta\epsilon^*= m\epsilon_\mathrm{T}/\mu_\mathrm{T} +\epsilon_\mathrm{F}$. 
The latter behavior has been clear observed in experiments~\cite{TrionExperiment1,TrionExperiment6}.  

The dependence of the amplitudes of trion and exciton peaks on the Fermi energy $\epsilon_\mathrm{F}$ are presented in Fig.~6-b. 
The exciton peak strength declines rapidly with increasing carrier density.  
The height of the trion peak depends more weakly on doping because of compensation spectral weight flow between
 polaronic branches in $A_\mathrm{X}(\omega,0)$ and the decrease of the exciton matrix element $D$ with doping. 
 The total spectral weight for the excitonic contribution to the optical conductivity
 is equal to $Z_\mathrm{\sigma}=2\pi \sigma_0 \bar{\epsilon}_\mathrm{X} D^2$ 
 and decreases as $D^2$ (see Fig.2-b), in agreement with 
 experiment~\cite{TrionExperiment1,TrionExperiment6,TMDCDEmler}. The spectral weights of peaks become comparable with each other and compete at $\epsilon_\mathrm{F}\sim\epsilon_\mathrm{T}$, where exciton-polaron picture is relevant.
  
The dependence of the widths of trion $\delta \epsilon_\mathrm{T}$ and exciton $\delta \epsilon_\mathrm{X}$ peaks (HWHM) on 
Fermi energy $\epsilon_\mathrm{F}$ are presented in Fig. 6-c. 
The width of the trion peak is doping independent and is equal to $2\gamma$, whereas the 
width of the exciton peak grows linearly with $\epsilon_\mathrm{F}$ as a result of scattering from Fermi sea fluctuations. 

Finally, we estimate the density range, where the exciton-polaron picture is relevant. For $\hbox{Mo}\hbox{S}_2$ with $m\approx 0.35\; m_0$, where $m_0$ is bare electronic mass, and  $\epsilon_\mathrm{T}=18\,\hbox{meV}$ we get the density range $n_\mathrm{e}\sim10^{12}\sim10^{13}\;\hbox{cm}^{-2}$. For $\hbox{CdTe}$ quantum wells with $m\approx 0.15 \;m_0$  and  $\epsilon_\mathrm{T}=2.1\,\hbox{meV}$ we get electronic density range $n_\mathrm{e}\sim10^{11}\;\hbox{cm}^{-2}$.

\section{VI. Conclusions}
We have developed a microscopic theory of absorption for moderately doped two-dimensional semiconductors. The theory takes into account both static and dynamical effects of Fermi sea formed by excess charge carriers. Static effects of Fermi sea renormalize energy of excitons and their coupling with light.   Dynamical excitations of the Fermi sea dress excitons into exciton-polarons, which are many-body generalization of trion bound and unbound states. As a result excitonic states split into attractive and repulsive exciton-polaron branches, which manifest themselves as two peaks in absorption. The calculated doping dependence of absorption is in good agreement with experiments.

We argue that, contrary to the conventional interpretation, the splitting can non been explained as a result of trions, weakly bound three-particle complexes. We have shown that in the  density range, where three-particle physics is involved, the trion feature is much smaller than one of excitons. In the density range, where they are comparable and compete with each other, exciton-polaron picture is appropriate.

\section{Acknowledgment} 
This material is upon work supported by the Army Research Office under Award No. W911NF-15-1-0466 and
by the Welch Foundation under Grant No. F1473. D.K.E is grateful to Fengcheng Wu for valuable
discussions. 

\bibliography{ExcitonPolaronBib}

\begin{thebibliography}{52}
\bibitem [{CC2()}]{CC2}%
  \BibitemOpen
  \href@noop {} {}\bibinfo {note} {The model we consider is the non-relatvistic
  limit of a four band model for massive Dirac particles with strong
  spin-valley coupling. The role of Berry phases and the Dirac-like spectrum of
  TMDC semiconductors has been discussed in several recent papers
  ~\cite{ExcitonDirac1,ExcitonDirac2,ExcitonDirac3,ExcitonDirac4,ExcitonDirac5,
  ExcitonDirac6}. They are important only for excited excitonic states, whch we
  exclude from consideration below.}\BibitemShut {Stop}%
\bibitem [{CC6()}]{CC6}%
  \BibitemOpen
  \href@noop {} {}\bibinfo {note} {A TMDC monolayer is usually surrounded by
  dielectric matter with different permittivity $\kappa'$, resulting in
  modification of bare Coulomb interactions $V_\vec{q}^0 F(q d/2)$, where $d$
  is TMDC effective thickness and $F(x)$ is given by~\cite{Keldysh}
  $$F(x)=\frac{\kappa^2-\kappa'^2+(\kappa^2+\kappa'^2)\mathrm{cosh}(x)+2 \kappa
  \kappa' \mathrm{sinh}(x)}{(\kappa^2+\kappa'^2)\mathrm{sinh}(x)+2 \kappa
  \kappa' \mathrm{cosh}(x)}.$$ For the ground excitonic state, which is only
  relevant in this work, $d\gg a_\mathrm{X}$ resulting in $F(0)\approx 1$ and
  the surrounding media is unimportant.}\BibitemShut {Stop}%
\bibitem [{CC4()}]{CC4}%
  \BibitemOpen
  \href@noop {} {}\bibinfo {note} {These relationships between $v$, $m$ and
  $\epsilon_\mathrm{g}$ are valid only in the non-relativistic limit of the
  simplified four-band Dirac model we are employing. Our results apply for more
  general underlying band models when $v$, $m$ and $\epsilon_\mathrm{g}$ are
  viewed as independent parameters.}\BibitemShut {Stop}%
\bibitem [{CC5()}]{CC5}%
  \BibitemOpen
  \href@noop {} {}\bibinfo {note} {The $^*$ accents here emphasize that the
  binding energies are renormalized in a non-trivial way at finite Fermi energy
  $\epsilon_\mathrm{F}$.}\BibitemShut {Stop}%
\end{thebibliography}

\begin{comment}

%%%%%%%%%% Merge with supplemental materials %%%%%%%%%%
\widetext
\clearpage
\begin{center}
\textbf{\large Supplemental Material: "Many-Body Theory of Trion Absorption Features in Two-Dimensional Semiconductor" by Dmitry K. Efimkin, and Allan H. MacDonald} 
\end{center}

%%%%%%%%%% Merge with supplemental materials %%%%%%%%%%
%%%%%%%%%% Prefix a "S" to all equations, figures, tables and reset the counter %%%%%%%%%%
\setcounter{equation}{0}
\setcounter{figure}{0}
\setcounter{table}{0}
\setcounter{page}{1}
\makeatletter
\renewcommand{\theequation}{S\arabic{equation}}
\renewcommand{\thefigure}{S\arabic{figure}}
\renewcommand{\bibnumfmt}[1]{[S#1]}
\renewcommand{\citenumfont}[1]{S#1}
%%%%%%%%%% Prefix a "S" to all equations, figures, tables and reset the counter %%%%%%%%%%

%\bibliographySM{bib}

\end{comment}

\begin{widetext}
	
\section{Appendix A. Excitonic contribution to optical conductivity}
Here we present detailed derivation of excitonic contribution to optical conductivity of a semiconductor. Real part of the optical conductivity $\sigma(\omega)$, which is responsible for the absorption, is connected with the retarded current-current response function $\chi^\mathrm{R}(\omega)$ as follows  $\sigma(\omega)=\Im[\chi^\mathrm{R}(\omega)]/\omega$. Excitons correspond to the ladder series of scattering diagrams and their summation can be reduced to the renormalization of the current vertex $W^0_\vec{p}\rightarrow W_\vec{p}$, as it depicted in Fig.~7-a and -b. We also take into account renormalization of the gap between conduction and valence bands due to Coulomb interactions in the Hartree-Fock approximation, as it is presented in Fig.~7-c and -d. The resulting current-current response function can be written as follows
\begin{equation}
\label{ChiM}
\chi(i \omega_n)= g_\mathrm{\alpha}T \sum_{\vec{p} p_n}\left[ W_\vec{p}(i \omega_n) G_\mathrm{c}(\bm{i}\omega_n + i p_n,\vec{p}) G_\mathrm{v}(\bm{i}p_n, \vec{k} ) W^0_\vec{p} + W^0_\vec{p} G_\mathrm{v}(\bm{i}\omega_n + i p_n,\vec{k}) G_\mathrm{c}(\bm{i}p_n, \vec{p} ) W_\vec{p}(-i \omega_n)\right],   
\end{equation}
where $w_n= 2 n\pi $ and $p_n= (2 n+1)\pi$ are bosonic and fermionic Matsubara frequencies. $g_\alpha$ is the degeneracy factor.  $W_\vec{p}^0=e v$ is the bare current vertex with $v=(\epsilon_\mathrm{g}/2m)^{1/2}$ to be a matrix element of velocity operator between conduction and valence bands. The renormalized current vertex $W_\vec{p}(i  \omega_n)$ satisfies the following integral equation 
\begin{equation}
\label{WM}
W_\vec{p}(i\omega_n)=W^0_\vec{p}+\sum_{\vec{p}' p_n} V_{\vec{p}-\vec{p}'} G_\mathrm{c}(\bm{i}\omega_n + i p_n,\vec{p}')G_\mathrm{v}( i p_n,\vec{p}')W_{\vec{p}'}(i \omega_n). 
\end{equation}
Electronic Green functions in (\ref{ChiM}) and (\ref{WM}) are given by $G_\mathrm{c}(i p_n,p)=(i p_n-\epsilon_\vec{p}^\mathrm{c}-\Sigma^\mathrm{c})$ and  $G_\mathrm{v}(i p_n,p)=(i p_n-\epsilon_\vec{p}^\mathrm{v}-\Sigma^\mathrm{v})$, where we have taken into account that for static interactions self-energies $\Sigma^\mathrm{c(v)}$ are frequency independent and neglect their momentum dependence implying $\Sigma^\mathrm{c(v)}=\Sigma^\mathrm{c(v)}_{\vec{p}=0}$. Physically, it means that we neglect the renormalization of electron masses in conduction and valence bands, but consider the renormalization of the gap  $\Sigma_\mathrm{g}=\Sigma^\mathrm{c}-\Sigma^\mathrm{v}$ between them. The latter can be presented as follows
\begin{equation}
\label{SigmaGapS}
\Sigma_\mathrm{g}=-\sum_{\vec{p}} V_\vec{p} n_\mathrm{F}(\epsilon_\vec{p}^\mathrm{c})- \sum_{\vec{p}}(V_\vec{p}^0-V_\vec{p})n_\mathrm{F}(\epsilon_\vec{p}^\mathrm{v}),
\end{equation}
where the first term is the exchange energy of an electron in the conduction band, while the second term is the modification of exchange energy of an electron in the valence band. Note that Hartree terms for electrons in conduction and valence bands exactly compensate each other, and $\Sigma_\mathrm{g}$ vanishes in the absence of doping $\epsilon_\mathrm{F}$.  

\begin{figure}[t]
	\label{S1}
	\vspace{-2 pt}
	\includegraphics[trim=6cm 12cm 2cm 1.2cm, clip,width=15 cm]{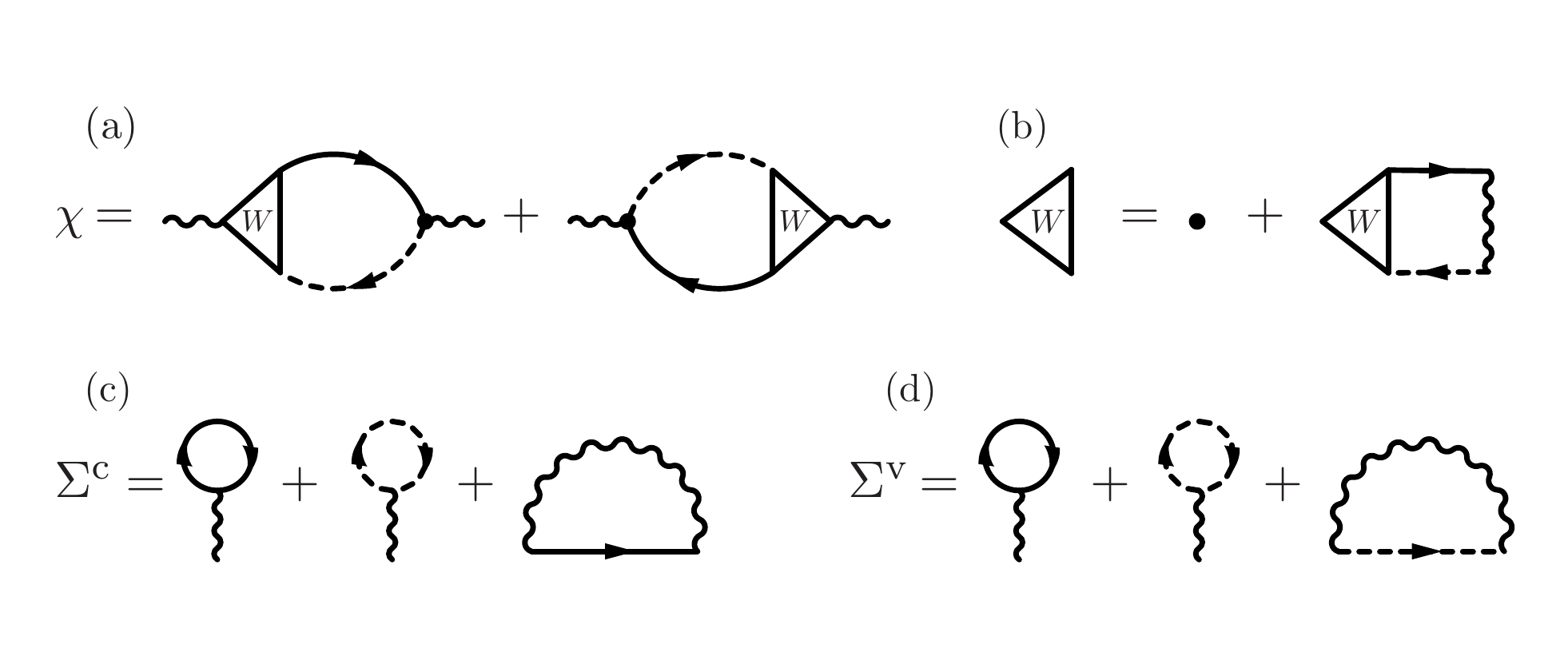}
	\vspace{-2 pt}
	\caption{(a) Diagrammatic representation for the current-current correlation function $\chi(i \omega_n)$. Solid and dashed lines correspond to electrons from conduction and valence bands. Plain (b) Excitons correspond to the ladder series of scattering diagrams and their summation can be reduced to the renormalization of current vertex $W_\vec{p}^0\rightarrow W_\vec{p}$. (c) and (d) self-energies of electrons in conduction and valence bands in the Hartree-Fock approximation. Hartee contributions (the first two terms) in $\Sigma^\mathrm{c}_\vec{p}$ and $\Sigma^\mathrm{v}_\vec{p}$ are equal to each other, renormalize chemical potential and do not influence the gap $\epsilon_\mathrm{g}$ between conduction and valence bands. The gap renormalization $\Sigma_\mathrm{g}=\Sigma^\mathrm{c}_{\vec{p}=0}-\Sigma^\mathrm{v}_{\vec{p}=0}\equiv\Sigma^\mathrm{c}-\Sigma^\mathrm{v}$ is governed by the difference of Fock terms representing exchange interactions between electrons.}
\end{figure}

After summation over Matsubara frequencies and analytical continuation $i \omega_n\rightarrow \omega+ i \gamma$, with $\gamma$ to be phenomenologically introduced decay rate of excitons, equations (\ref{ChiM}) and (\ref{WM}) reduce to 
\begin{equation}
\label{ChiR1}
\chi^\mathrm{R}(\omega)=-g_\mathrm{\alpha}\sum_{\vec{p}} \left[ b_\vec{p} W_\vec{p}' (\omega) W^0_\vec{p}  + b_\vec{p} W^0_\vec{p} W_\vec{p}' (-\omega)\right].
\end{equation}
\vspace{-0.4 cm} 
\begin{equation}
\label{WR}
\left[\frac{p^2}{2 \mu_\mathrm{X}}+\Sigma_\mathrm{g} \right] W'_\vec{p}(\omega) - \sum_{\vec{p}'} b_\vec{p} V_{\vec{p}-\vec{p}'} b_{\vec{p}'} W'_{\vec{p}'}(\omega) + b_\vec{p} W_\vec{p}^0= (\omega-\epsilon_\mathrm{g}+i \gamma) W'_\vec{p}(\omega).   
\end{equation}
Here we have introduced  $W'_\vec{p}(\omega)=b_\vec{p}W^\mathrm{R}_\vec{p}(\omega)/(\omega-\vec{p}^2/2\mu_\mathrm{X}-\epsilon_\mathrm{g}-\Sigma_\mathrm{g}+i \gamma)$ with reduced mass of electron and hole, $\mu_\mathrm{X}=m/2$. It is instructive to introduce the auxiliary eigenvalue problem, which represents Schroedinger-like equation in the momentum space, as follows   
\begin{equation}
\left[\frac{\vec{p}^2}{2 \mu_\mathrm{X}}+\Sigma_\mathrm{g}\right] C_\vec{p} - \sum_{\vec{p}'} b_\vec{p} V_{\vec{p}-\vec{p}'} b_{\vec{p}'}C_{\vec{p}'}=\epsilon_\mathrm{X} C_\vec{p} .
\end{equation}
Here $\epsilon_X$ is a binding energy of an exciton, while $C_\vec{p} $ is its wave function in the momentum space. Due to the rotational symmetry of the problem, the eigenvalues can be numbered by main $n$ and orbital $m$ quantum numbers. With the normalization condition $\sum_{\vec{p}}|C_\vec{p}|^2=1$, they form the complete set of states,  which can be used for a decomposition of $W_\vec{p}'$  as follows $W_\vec{p}'=\sum_{nm} W'_{nm}C_\vec{p}^{nm}$. Its substitution in (\ref{WR}), and integration over momentum results in 
\begin{equation}
\label{ME}
W'_{nm}(\omega)= e \sqrt{\frac{\epsilon_\mathrm{g}\epsilon_\mathrm{X}}{\pi}} \frac{D_{nm}^*}{\omega-\epsilon_\mathrm{g}-\epsilon_\mathrm{X}^{nm}+i \gamma}, \quad \quad D_{nm}=\sqrt{\frac{\pi }{2 \bar{p}_\mathrm{X}^2}} \sum_{\vec{p}} b_\vec{p} C_\vec{p}^{nm}.
\end{equation}
Here $D_{nm}$ is the dimensionless matrix element for exciton-light coupling and we have introduced  $\bar{\epsilon}_\mathrm{X}=m e^4/\kappa^2 \hbar^2$ along with  $\bar{p}_\mathrm{X}=me^2/\kappa \hbar$. They are the binding energy and the stretch of wave function in the momentum space for the ground excitonic state in the absence of doping. Substitution of (\ref{ME}) to (\ref{ChiR1}) results in
\begin{equation}
\label{ChiR2}
\frac{
	\chi^\mathrm{R}(\omega)}{\sigma_0} =-g_\mathrm{\alpha} \epsilon_\mathrm{g}  \sum_{nm} |D^{nm}|^2\left[ \frac{2 \bar{\epsilon}_X}{\omega-\epsilon_\mathrm{g}-\epsilon_\mathrm{X}^{nm}+i \gamma} + \frac{2 \bar{\epsilon}_X }{-\omega-\epsilon_\mathrm{g}-\epsilon_\mathrm{X}^{nm}+i \gamma } \right],
\end{equation}
where $\sigma_0=e^2/h$ is the conductivity quanta. Recalling that $\sigma(\omega)=\Im[\chi^\mathrm{R}(\omega)]/\omega$ and taking into account that  $\epsilon_\mathrm{X}^{nm}\ll\epsilon_\mathrm{g}$ we get 
\begin{equation}
\frac{\sigma(\omega)}{\sigma_0}=g_\mathrm{\alpha}\sum_{nm} |D^{nm}|^2 \left[\bar{\epsilon}_\mathrm{X} A_\mathrm{X} (\omega,0) + \bar{\epsilon}_\mathrm{X} A_\mathrm{X} (-\omega,0)\right].
\end{equation}
Here we have introduced the spectral function of excitons $A_\mathrm{X}(\omega,\vec{q})=-2 \Im[G_\mathrm{X}(w,\vec{q})]$ and their function is given by $G_\mathrm{X}^\mathrm{R}(\omega,\vec{q})=(\omega-\epsilon_\mathrm{g}-\epsilon_\mathrm{X}-\vec{p}^2/2 M_\mathrm{T}+i \gamma)^{-1}$ with excitonic mass $M_\mathrm{T}=2m$. Note that the real part of optical conductivity  $\sigma(\omega)=\sigma(-\omega)$ is an even function of frequency, which is a general property of the dissipative part of response functions~\cite{MahanSM}. Without loss of generality, we can restrict $\omega$ only to positive frequencies and measure it from the gap, $\omega\rightarrow \omega + \epsilon_\mathrm{g}$, as we do in the paper. As a result, we get Eq.~(4) from the paper.

\section{Appendix B. Interactions between exciton and electron}
In the paper we introduce attractive interactions between exciton and electron $U$ in a phenomenological way and treat it as an independent parameter of our theory. Here we present estimations of $U$ and the binding energy for electron and exciton $\epsilon_\mathrm{T}$. 

The attraction between an exciton and an electron appears due to the polarization mechanism. An exciton is polarized by electric field of an electron with magnitude $E=e/\kappa R^2$, where $R$ is distance between them, acquires a dipole moment $\vec{p}=\alpha \vec{E}$, where $\alpha$ is exciton polarizability, and gets potential energy 
\begin{equation}
V(R)=-\frac{\alpha \vec{E}^2}{2}=-\frac{\alpha e^2}{2 \kappa ^2 R^4}
\end{equation}
To calculate the polarizability of the exciton $\alpha$ we use quantum mechanical perturbation theory. Interaction energy with electric field $\vec{E}$, which we treat as a perturbation is, $H_\mathrm{E}=-e \vec{r} \vec{E}$, where $\vec{r}$ is the relative distance between electron and hole. Exciton is assumed to be in the ground state $|n=0,m=0\rangle$, and due to its $s$-wave nature the first order correction to the energy is zero, $V_1=\langle 0,0| H_\mathrm{E}|0,0\rangle=0$. The second order term can be written as follows
\begin{equation}
\label{V2}
V_2=\sum_{nm} \frac{|\langle 0,0|H_\mathrm{E}|n,m\rangle|^2}{\epsilon_\mathrm{X}^{00}-\epsilon_\mathrm{X}^{nm}} + \sum_{\vec{p}} \frac{|\langle 0,0|H_\mathrm{E}|\vec{p}\rangle|^2}{\epsilon_\mathrm{X}^{00}-\epsilon_\mathrm{X}^\vec{p}}.
\end{equation}  
The first term describes virtual transitions from the ground to excited localized states, while the second one describes virtual ionization transitions. In the doped regime, we consider in the paper, excited states merge with continuum and only the second term in (\ref{V2}) survives. For estimations we use the ground state wave function in the absence of doping, and approximate delocalized states by plane waves as following 
\begin{equation}
\label{SetC}
C^{00}_\vec{r}=\frac{2}{\bar{a}_\mathrm{X}} e^{-r/\bar{a}_\mathrm{X}}, \qquad \epsilon_\mathrm{X}^{00}=\Sigma_\mathrm{g} -\bar{\epsilon}_\mathrm{X}  \qquad \qquad \hbox{and} \qquad \qquad  C^\vec{k}_\vec{r}=\frac{1}{\sqrt{S}} e^{i \vec{p} \vec{r}/\hbar}, \qquad  \epsilon_\mathrm{X}^\vec{p}=\Sigma_\mathrm{g} + \frac{\vec{p}^2}{2\mu_\mathrm{X}}.			 
\end{equation}
where $\bar{a}_\mathrm{X}=\hbar \kappa/m e^2$ and $\bar{\epsilon}_X=m e^4/\hbar \kappa^2$ are radius and binding energy of the excitons. $S$ is the area of  considered two-dimensional system. We measure energies from the bottom of the conduction band in the absence of doping as we do in the paper. $\Sigma_\mathrm{g}$ is the gap renormalization, which is completely unimportant here since only difference between energies is involved in (\ref{V2}). The set of wave functions (\ref{SetC}) results in the following matrix element
\begin{equation}
\label{MatrixElementSM} 
\langle 0,0|H_\mathrm{E}|\vec{p}\rangle=-e\vec{p} \vec{E} \frac{12 \pi \bar{a}_\mathrm{X}^3}{\sqrt{S}} \frac{1}{[1+(p \bar{a}_\mathrm{X})^2]^{5/2}}.
\end{equation}
After substitution of (\ref{MatrixElementSM}) to (\ref{V2}) we get
\begin{equation}
V_2=-\frac{\alpha \vec{E}^2}{2}, \quad\quad \quad \alpha=\frac{8}{5} \frac{e^2 \bar{a}_\mathrm{X}^2}{\bar{\epsilon}_\mathrm{X}}.
\end{equation}

Interaction constant $U$ correspond to the Fourie transform $V(\vec{q}=0)$ at zero momenta. The latter is diverging and we regularize the interactions at the excitonic radius as follows $V_\mathrm{reg}(\vec{R})=-\alpha e^2/2 \kappa^2 (R^2+\bar{a}_\mathrm{X}^2)^2$, which results in $U=V_\mathrm{reg}(\vec{q}=0)=\pi \alpha e^2/2\kappa^2 \bar{a}_\mathrm{X}^2=16\pi \bar{\epsilon}_\mathrm{X} a^2/5$.

The binding energy of trion is given by $\epsilon=\hbar^2/2\mu_\mathrm{T} \bar{a}_\mathrm{X}^2 \times \exp[-2\pi \hbar^2/\mu_\mathrm{T} U]=3\bar{\epsilon}_\mathrm{X}/4 \times \exp[-15/16]$, where $\mu_\mathrm{T}=2 m/3$ is reduced mass of exciton and electron, and we take the momentum cutoff $p_\mathrm{\Lambda}=\hbar/\bar{a}_\mathrm{X}$. As a result we get $\epsilon_\mathrm{T}/\epsilon_\mathrm{X}\approx0.3$, which overestimates their ration in experiments $\epsilon_\mathrm{T}/\epsilon_\mathrm{X}\approx0.07$. It should be noted that the estimations for $\epsilon_\mathrm{T}$ are quite sensitive to the cutoff and the regularization procedure, hence they are supposed to give only the correct order of magnitude. 

\begin{figure}[t]
	\label{FigS2}
	\includegraphics[width=9.6 cm]{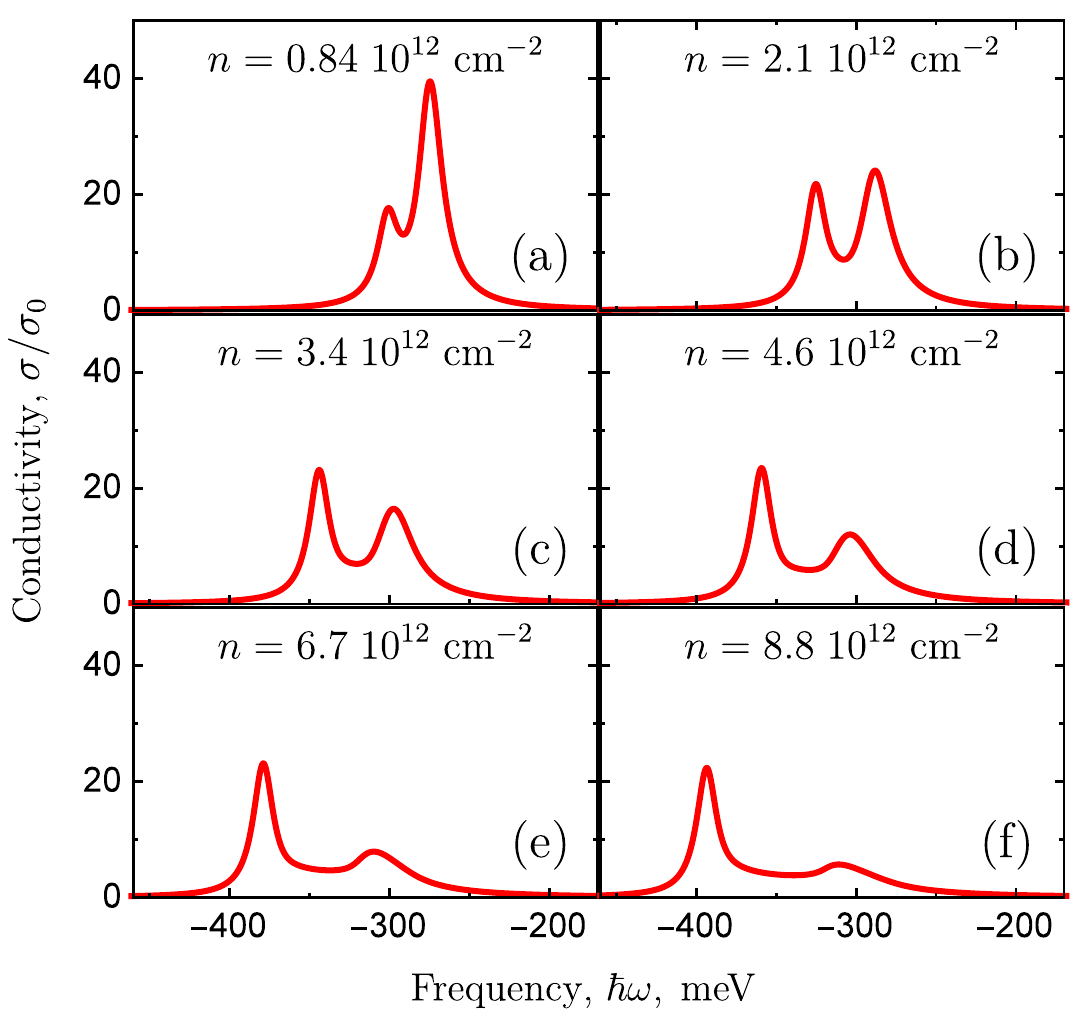} \qquad
	\includegraphics[width=6.6 cm]{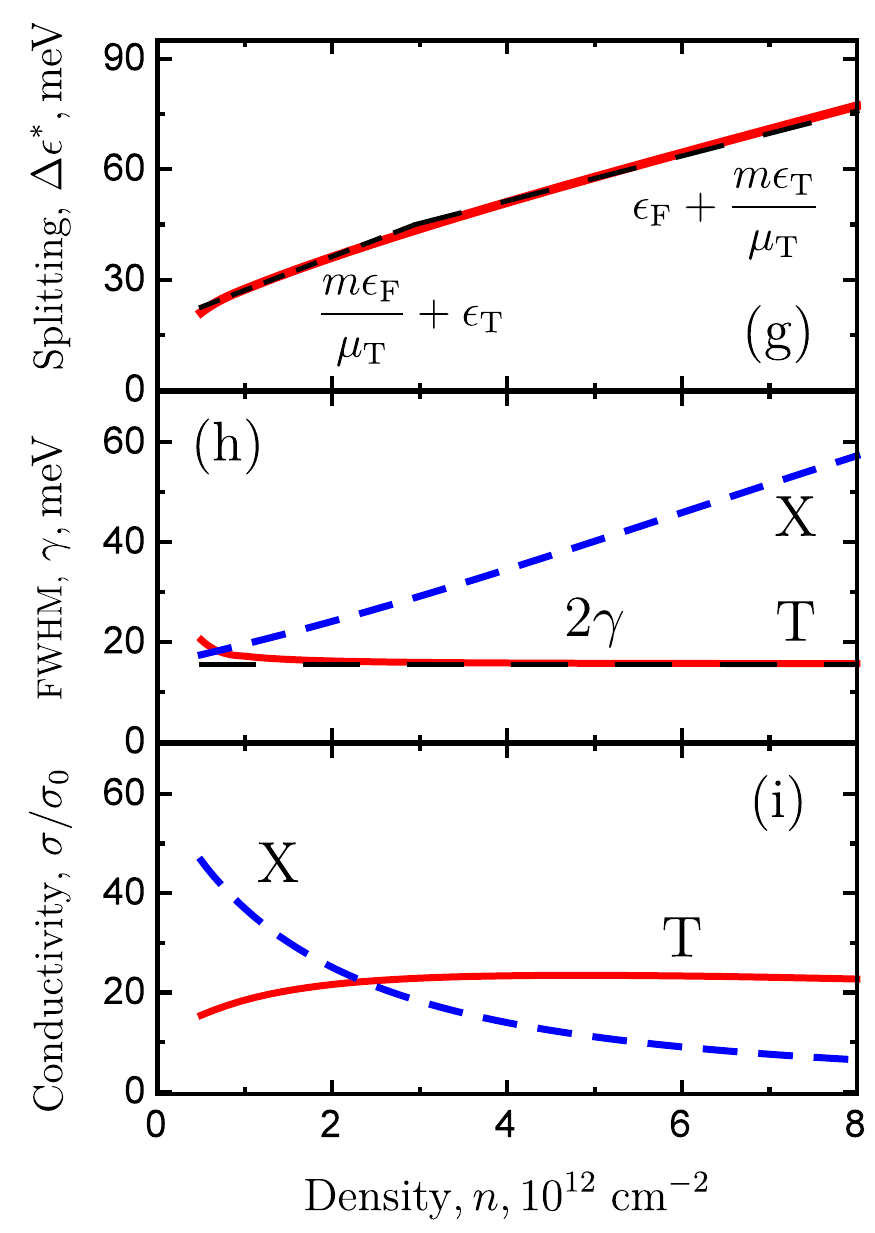}
	\vspace{-2 pt}
	\caption{$\hbox{(a)}$-$\hbox{(f)}$ Frequency dependence of the optical conductivity $\sigma(\omega)$ for different values of electron density $n$. $\hbox{(h)}$ Density dependence of the energy splitting $\Delta \epsilon^*=\epsilon_\mathrm{X}^*-\epsilon_\mathrm{T}^*$ between exciton $\hbox{X}$ and trion $\hbox{T}$ absorption features. Density dependence of absorption feature $\hbox{(h)}$  peaks $\sigma(\epsilon_\mathrm{X(T)}^*)/\sigma_0$  and
		$\hbox{(i)}$ widths $\delta \epsilon_\mathrm{X(T)}/\bar{\epsilon}_\mathrm{X}$. The splitting interpolates between two linear behaviors $\Delta\epsilon^*= \epsilon_\mathrm{T}+m\epsilon_\mathrm{F}/\mu_\mathrm{T}$ at $\epsilon_\mathrm{F}\lesssim \epsilon_\mathrm{T}$, and $\Delta\epsilon^*= m\epsilon_\mathrm{T}/\mu_\mathrm{T} +\epsilon_\mathrm{F}$ at $\epsilon_\mathrm{F}\gtrsim \epsilon_\mathrm{T}$. }
\end{figure}

\section{Appendix C. Plots in real units}
In the main text of the paper we present results in dimensionless units. Here we replot Fig.5 and Fig.6 in real units. For calculations we have used the set of parameters $\epsilon_\mathrm{T}=18\; \hbox{meV}$, $\epsilon_\mathrm{X}\approx 260\;\hbox{meV}$, $m=0.35\; m_0$, where $m_0$ is the bare mass of electrons, relevant to $\hbox{MoS}_2$. Density dependence of absorption is presented in Fig.~8.

\section{Appendix D. Spectral weight of trions $Z_\mathrm{T}$}
Here we present an analytical calculation of the spectral weight of trions $Z_\mathrm{T}$ in the low-density regime $\epsilon_\mathrm{F}\ll\epsilon_\mathrm{T}$, where the exciton-electron problem reduces to  two-particle one. In that regime $\Gamma^\mathrm{R}(\omega,\vec{q})$ reduces to the exact two-particle $T$-matrix, given by
\begin{equation}
\Gamma^\mathrm{R}(\omega,\vec{q})=\frac{2\pi \hbar^2}{\mu_\mathrm{T}}\frac{1}{\log\left[\frac{\epsilon_\mathrm{T}}{\omega-\vec{q}^2/2M_\mathrm{T}-\epsilon_\mathrm{X}+ i\gamma}\right]+\bm{i} \pi},
\end{equation}
As a result, the self-energy of excitons $\Sigma_\mathrm{X}(\omega,0)$ can be approximated as follows
\begin{equation}
\Sigma_\mathrm{X}^\mathrm{R}(\omega,0)=\sum_{\vec{p}} \Gamma^\mathrm{R}(\omega+\epsilon_{\vec{p}}^\mathrm{c},\vec{p})n_\mathrm{F}(\epsilon_{\vec{p}}^\mathrm{c})\approx  \frac{\Sigma_0}{\log\left[\frac{\epsilon_\mathrm{T}}{\omega-\epsilon_\mathrm{X}+i \gamma}\right]+\bm{i} \pi},
\label{SelfEnergyS}
\end{equation} 
where $\Sigma_0=\epsilon_\mathrm{F} m/\mu_\mathrm{T}$. The self-energy defines spectral function of excitons $A_\mathrm{X}(\omega,0)=-2 \Im[\left\{\omega-\epsilon_\mathrm{X}-\Sigma(\omega,0)\right\}^{-1}]$. Solutions of the equation $\omega^*-\epsilon_\mathrm{X}-\Re[\Sigma^R(\omega^*,0)]=0$ correspond to quasiparticle peaks in $A_\mathrm{X}(\omega,0)$. In the absence of doping,  the spectral function of excitons has the only peak at $\epsilon_\mathrm{X}^*=\epsilon_\mathrm{X}$ corresponding to bare excitons.  In the low doping regime the self-energy is small $\Sigma^\mathrm{R}(\omega,0)/\epsilon_\mathrm{T}\sim\epsilon_\mathrm{F}/\epsilon_\mathrm{T}\ll 1$ at all frequencies except vicinity of singularity at $\omega=\epsilon_\mathrm{X}-\epsilon_\mathrm{T}$, which appears due to the presence of the exciton-electron bound state pole in $\Gamma^\mathrm{R}(\omega,\vec{q})$. In the vicinity of the singularity the self-energy is given by 
\begin{equation}
\Sigma_\mathrm{X}^\mathrm{R}(\omega,0)\approx\frac{\Sigma_0 \epsilon_\mathrm{T}}{\omega-\epsilon_\mathrm{X}+\epsilon_\mathrm{T}}-i  \frac{\gamma \Sigma_0  \epsilon_\mathrm{T}}{\gamma^2+(\omega-\epsilon_\mathrm{X}+\epsilon_\mathrm{T})^2}.
\end{equation}
The presence of the singularity leads to an additional trion peak in the spectral function of excitons $A_\mathrm{X}(\omega,0)$ at energy   $\epsilon_\mathrm{T}^*\approx\epsilon_\mathrm{X}-\epsilon_\mathrm{T}-\Sigma_0$, while the position of the exciton peak is weakly modified $\epsilon_X^*\approx \epsilon_X$, since $\Sigma^\mathrm{R}(\epsilon_\mathrm{X},0)/\epsilon_\mathrm{T}\sim \epsilon_\mathrm{F}/\epsilon_\mathrm{T}\ll 1$. In the vicinity of trion peak the spectral function is given by  
\begin{equation}
A_\mathrm{X}^\mathrm{T}(\omega,0)\approx Z_\mathrm{T} \frac{2 \gamma_\mathrm{T}}{(\omega-\epsilon_\mathrm{T}^*)^2+\gamma_\mathrm{T}^2}\approx 2\pi Z_\mathrm{T} \delta(\omega-\epsilon_\mathrm{T}^*),
\end{equation}
where $Z_\mathrm{T}=\Sigma_0/\epsilon_\mathrm{T}$ is the spectral weight of trions and $\gamma_\mathrm{T}=\gamma \Sigma_0/\epsilon_\mathrm{T}+\gamma\Sigma_0^2/(\gamma^2+\Sigma_0^2)$ is their decay rate. The last equality implies $\gamma_\mathrm{T}\ll \epsilon_\mathrm{T}$, which is satisfied at $\Sigma_0/\epsilon_\mathrm{T}\ll 1$ and $\gamma/\epsilon_\mathrm{T}\lesssim 1$. Since the total spectral weight is conserved the spectral function of excitons in the low density regime can be approximated as follows 
\begin{equation}
A_\mathrm{X}(\omega,0)\approx 2\pi Z_\mathrm{T} \delta(\omega-\epsilon_\mathrm{T}^*)+2\pi (1-Z_\mathrm{T}) \delta(\omega-\epsilon_\mathrm{X}^*).
\end{equation}
Note that the spectral weight of trions is much smaller than one of excitons and splitting between peaks $\Delta \epsilon^*=\epsilon_\mathrm{X}^*-\epsilon_\mathrm{T}^*=\epsilon_\mathrm{T}+\epsilon_\mathrm{F}m/\mu_\mathrm{T}$ goes linearly with Fermi energy $\epsilon_\mathrm{F}$.
%\end{comment}

\end{widetext}
\end{document}